# Survey and Improvement Strategies for Gene Prioritization with Large Language Models

Matthew B. Neeley, B.S.,[1,2]* Guantong Qi, B.S.,[1,3]* Guanchu Wang, M.S.,[4]* Ruixiang Tang, PhD,[5] Dongxue Mao, Ph.D.,[1,6,7] Chaozhong Liu, Ph.D.,[1,2] Sasidhar Pasupuleti, M.S.,[1,6] Bo Yuan, Ph.D.,[7,8] Fan Xia, Ph.D.,[7,9] Pengfei Liu, Ph.D.,[7,9] Zhandong Liu, Ph.D.,[1,6]** Xia Hu, Ph.D.,[4]**

Affiliations:
Co-first authors *, Co-corresponding authors **, [1] Jan and Dan Duncan Neurological Research Institute at Texas Children's Hospital, Houston, [2] Graduate School of Biomedical Sciences, Program in Quantitative and Computational Biosciences, Baylor College of Medicine, Houston, [3] Graduate School of Biomedical Sciences, Program in Genetics and Genomics, Baylor College of Medicine, Houston, [4] Department of Computer Science, Rice University, Houston, [5] Department of Computer Science, Rutgers University, New Jersey, [6] Department of Pediatrics, Baylor College of Medicine, Houston, [7] Department of Molecular and Human Genetics, Baylor College of Medicine, Houston, [8] Human Genome Sequencing Center, Baylor College of Medicine, Houston, [9] Baylor Genetics, Houston.

## Abstract

**Background**
Rare diseases can be difficult to diagnose due to limited patient data and broad genetic diversity. Despite the advances in variant prioritization tools, many rare disease cases remain undiagnosed. While large language models (LLMs) have performed well in medical exams, their accuracy in diagnosing rare genetic diseases has not yet been evaluated.

**Methods**
To identify causal genes for genetic diseases, we benchmarked various LLMs for gene prioritization. We employed multi-agent and Human Phenotype Ontology (HPO) classification approaches to identify patient case groups based on phenotypes categorizing them into different levels of solvability. To address LLM limitations in ranking large gene sets, we used a divide-and-conquer strategy to break down the ranking task into smaller subsets. Mini-batching inputs and limiting the number of generated tokens improved efficiency.

**Results**
In its vanilla form, GPT-4 consistently outperformed the other LLMs with an accuracy around 30%. Multi-agent and HPO classification approaches aided us in distinguishing between confidently-solved and challenging cases. In addition, we observed bias towards well-studied genes and input order sensitivity which are drawbacks of LLMs in disease-causal gene prioritization. Our divide-and-conquer strategy enhanced accuracy by overcoming positional and gene frequency biases in literature. This framework, based on benchmarking insights and novel techniques, significantly optimizes the overall process when identifying disease-causal genes for genetic diseases when compared to our baseline evaluation, and thereby better enables the development of targeted diagnostic and therapeutic interventions.

**Conclusions**
Using HPO classification and our novel multi-agent techniques, and our LLM divide-and-conquer strategy (1) highlighted the importance of accounting for differences in patient case solvability and (2) yielded improved performance in identifying causal genes for rare diseases when compared to our baseline evaluation. We anticipate that our approach will streamline the diagnosis of rare genetic disorders, facilitate the reanalysis of unsolved cases, and accelerate the discovery of novel disease-associated genes.

## Introduction

Rare diseases affect a small fraction of the population but collectively impose a significant burden upon individuals and society [1]. These rare conditions are frequently of genetic origin and classified as Mendelian diseases, and result from mutations in single genes. Their rarity and diversity considerably complicate diagnosis and treatment. A notable portion of these conditions remains undiagnosed due to the unpredictability and complexity of their genetic foundations and clinical presentations, leading to lifetimes of difficulty and uncertainty for patients and their families [2]. There is an urgent need for enhanced diagnostic tools and methodologies to evaluate patient data.

In response to this need, clinical sequencing, including whole genome sequencing (WGS) and exome sequencing (ES), has become an indispensable tool in the exploration of rare diseases. These techniques allow clinicians and researchers to meticulously analyze a patient's genome and detect potentially causative genetic variants. While the vast array of candidate variants presents a challenge, variant prioritization tools effectively address this issue.

Learning how to use gene prioritization to reliably identify the causal disrupted gene is critical for accurate diagnosis and understanding the genetic variants responsible for the disease [3,4]. Other approaches to gene prioritization depended largely on structured databases that compile well-known genetic data and predictions regarding the impact of various mutations [5,6]. However, due to the rarity and distinct complexity of rare diseases [7], traditional approaches to gene prioritization are less likely to be effective, leaving many rare-disease cases unresolved.

Innovative solutions in artificial intelligence have been primarily driven by the development of LLMs like GPT-4, Llama2[8], and Mixtral-8x7b[9]. These models are trained on a wide array of data sources, including textual datasets, programming code bases, and other forms of digital content. Such training enables these models to not only emulate human-like communication, but also to acquire a substantial understanding of specialized domains, including medical sciences [10,11,12,13]. Unlike simple data repositories, LLMs exhibit advanced reasoning capabilities, allowing them to process and interpret extensive and unstructured datasets effectively. This attribute is particularly valuable in exploring underutilized applications in genetic research and the diagnosis of rare diseases [14].

LLMs have been shown to excel at and achieve state-of-the-art performance in a variety of tasks and benchmarks [11,15,16]. Recently, in the medical setting, researchers found that LLMs outperform medical experts at summarizing electronic health records[17].

LLMs offer significant advantages in the rare-disease context, in which traditional data sources are scarce and unique [2]. LLMs excel at parsing structured and unstructured data and at converting clinical narratives and genetic information into structured data that enhances gene-based prioritization. Unlike traditional machine learning methods that require large datasets for training, LLMs can make a diagnosis prediction from a single case without task specific pre-training [16,18]. This capability allows LLMs to provide valuable insights and predictions even when working with limited and unique patient cases.

In this study, we explore the potential of LLMs to enhance gene prioritization by effectively analyzing clinical phenotypes and candidate disease-causing genes to identify the diagnostic gene. Utilizing LLMs allows us to refine both the accuracy of gene-based prioritization and the process by which we diagnose rare genetic disorders, ultimately leading to more precise and informed medical interventions.

## Methods

**Dataset**
Because there is not currently a consensus strategy for benchmarking gene-based prioritization methodology, synthetic datasets are often selected for evaluation [19]. We, however, used a large dataset of resolved patient cases: 1063 cases from a clinical diagnostic lab, Baylor Genetics (BG); 90 from the Undiagnosed Diseases Network (UDN); and 200 from Deciphering Developmental Disorders (DDD). The patient data consists of phenotypes from clinical electronic health records (EHR) and genomic sequencing data.

We preliminarily filtered raw genomic data from each patient to keep variants with an allele frequency less than 1%, or that were classified as pathogenic or likely pathogenic in ClinVar or HGMD, or that had a SpliceAI score greater than 0.8. The variants were converted to the gene level for gene-based prioritization. We then generated three gene sets of different sizes (5, 25, and 50 genes) for each patient by randomly sampling from their non-causal genes and adding the causal gene. We shuffled the genes to randomize to the gene ordering for the prompt. We finally converted patient clinical phenotypes from numerical HPO terms to their corresponding definitions because LLMs tend to hallucinate the true meaning of HPO terms (Fig. S10).

**Prompt Creation**
We enumerated clinical phenotypes and candidate genes to structure the prompt. Then we included instructions for the model to assign a rank based on association between the specified genes and the clinical phenotypes. We included further guidance regarding the use of gene function, expression sites, or analogous animal model information in the absence of direct empirical data. As a final point, to precisely define the response format, we instructed the model to rank gene associations by probability (0-1.0), emphasizing brevity and the exclusion of explanatory justifications (Fig. S1A).

**Multi-Agent Classification**
The Multi-Agent Classification method is an approach we developed for classifying patients by phenotypes via a multi-agent LLM pipeline that consists of two main steps (Fig. S4). In the first step, the original prompts are given to an evaluator agent (GPT-4) to write a 100-word essay that evaluates all the gene candidates in the prompt. In the second step, a summarizer agent summarizes the output essay and distinguishes whether there is at least one gene directly linked to the phenotypes by outputting "Yes" or "No". The "Yes" or "No" output for each case is then used to classify the cases into specific (Yes) or non-specific (No) groups for subsequent evaluation.

**HPO Phenotype Classification**
In this study, we adopt the HPO Phenotype Classification methodology (as detailed in the Cohort Analyzer tool publication [20]) to evaluate HPO term specificity regarding the HPO hierarchy termed the dataset specificity index (DsI). This metric assesses whether the HPO terms used to phenotype the patients are general (closer to the root node in the hierarchy) or specific (further from the root node). The DsI calculation considers the distribution of HPO terms across various levels of the HPO tree structure, penalizing general terms near the root while rewarding specific terms closer to the leaf nodes. For classification analyses, we randomly sampled 90 cases each from BG, UDN, and DDD for 270 total patient cases.

**Divide-and-Conquer Strategy**
LLMs struggle to effectively address problems involving numerous options [21]. To address this limitation, we used a divide-and-conquer strategy that consists of three steps: First, all gene candidates are randomly split into groups of five; second, LLMs estimate the probability of the gene candidates to be disease-causal within each group as their in-group probabilities; third, the final score for each gene is derived from the average of its in-group probability (Fig. S8). To formulate this pipeline, we give the final score $S(g_j)$ for a gene $g_j$ in the following equation:

$$S(g_j) = \frac{1}{N}\sum_{i=1}^{N} Pr(g_j)$$

Here, $N$ is the total sampling number and $Pr(g_j)$ denotes the in-group probability of the gene ( $g_j$ ) provided by the LLMs. We divided the 25 gene candidates into five groups and the 50 gene candidates into ten groups so that each group consists of 5 genes.

**Ranking Strategy**
We employ different strategies for interacting with proprietary and open-source LLMs due to their distinct capabilities. Specifically, GPT-3.5 and GPT-4 rank genes within their textual outputs. We selected to use a ranking strategy for open-source LLMs like Llama-2 and Mixtral-8x7B due to their ability to rank based on the probability of their output tokens, a feature not available with GPT-3.5 or GPT-4. This necessitates the use of a different prompt for open-source LLMs (Fig. S1B, Fig S2).

The open-source LLMs output either 'Yes' or 'No' to indicate the gene's $g_j$ causality to the phenotypes. We estimate a log-likelihood ratio of 'Yes' and 'No' [22] for determining the genes' causality to the phenotypes. The log-likelihood ratio is given as follows:

$$S(g_j) = log\left(\frac{Pr(Yes)}{Pr(No)}\right),$$

where $Pr(Yes)$ and $Pr(No)$ denote the output probability of the 'Yes' and 'No' tokens. The ranking of genes is determined based on their respective log-likelihood values.

# Results

### Benchmark of Large Language Models

We began our study by conducting a baseline evaluation on several LLMs as applied to three datasets of patients with rare diseases from Baylor Genetics, Deciphering Developmental Disorders, and the Undiagnosed Diseases Network. The LLMs we examined included GPT-4 [23], GPT-3.5 [23], Mixtral-8x7B [9], Llama-2-70B [8], and a specialist biomedical model called BioMistral [24]. These represent proprietary models, open-source models, a mixture of experts model, and a domain specialist model. For each patient case, the LLM evaluated lists of candidate genes with lengths of 5, 25, and 50, along with the patient's phenotypes, to observe its performance with different input sizes. The output was a ranked list of the genes with probabilities reflecting their likelihood of being associated with the phenotypes (Fig. 1A).

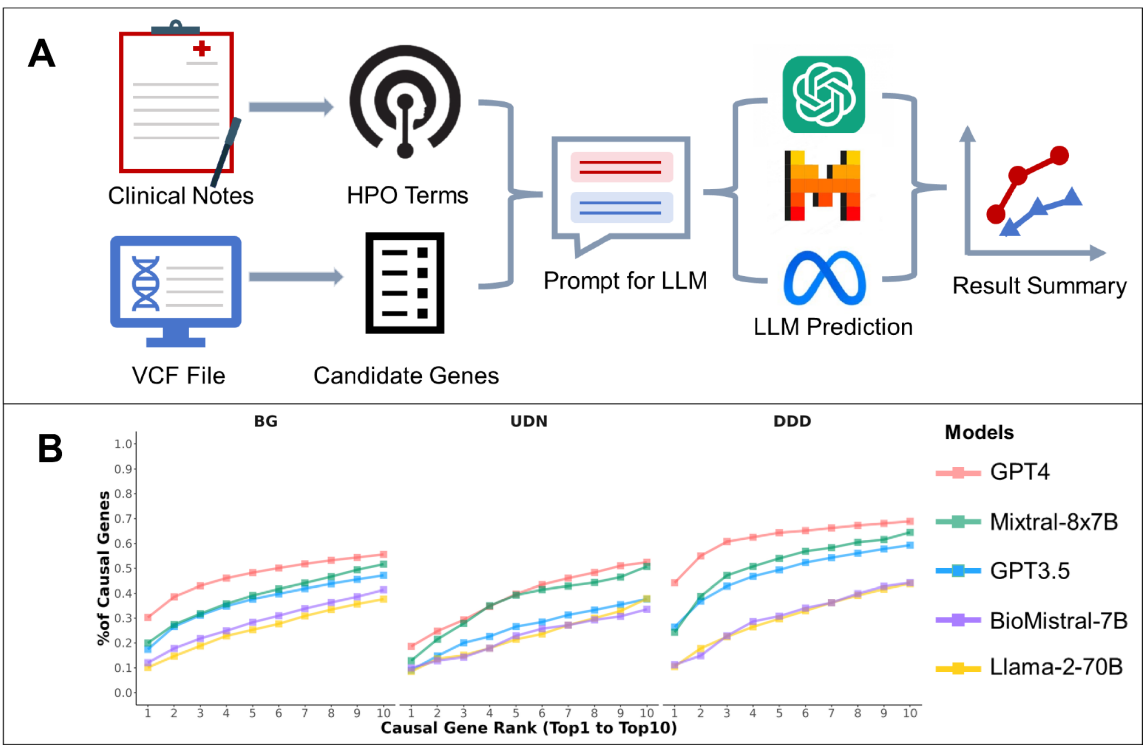

Figure 1. (Panel A) The workflow for prompt creation and baseline comparison of LLMs. (Panel B) Comparison of LLMs performance in ranking 50 causal genes. The x-axis represents the rank threshold, from 1 to 10, at or within which the presence of causal genes is evaluated. The y-axis shows the proportion of total causal genes captured at each rank threshold, calculated as the cumulative number of causal genes at or within a given rank across all patient cases, divided by the total number of causal genes. Higher values at lower rank thresholds indicate better performance in prioritizing causal genes. GPT-4 demonstrates the highest proportion of causal genes at the top ranks, followed by Mixtral-8x7B, GPT-3.5, BioMistral-7B, and Llama-2-7B.

Employing an ensemble prompting strategy [25], which averaged the results of five iterations for each prompt, we obtained a consensus result for each LLM. As is common practice, we also employed mini-batching and output clipping to enhance processing speeds for larger open-source models. GPT-4 emerged as the standout performer, demonstrating superior performance to the other LLMs in correctly ranking causal genes as the most likely candidates (ranked first) for the patient's disorder, highlighting its strong potential for use in genetic diagnostics (Fig. 1B, Fig. S3).

The performance gap between GPT-4 and the next best model, Mixtral-8x7B, is most pronounced in the DDD dataset, where GPT-4 captures 40% of causal genes at rank 1, compared to 20% for both GPT-3.5 and Mixtral-8x7B (Fig. 1B, Fig. S3). Across all datasets, Biomistral-7B and Llama-2-70B consistently exhibit low performance. Notably, identifying causal genetic genes proved more challenging for LLMs in the UDN dataset than in datasets from BG and DDD; this is likely because UDN patients tend to have had extensive prior screening without successful diagnosis or disorders not yet characterized, resulting in a wide range of phenotypes [26]. DDD cases target congenital or early-onset severe phenotypes in children, optimizing the identification of a highly penetrant monogenic cause [27]. BG consists of patients from a clinical diagnostic lab with varied cases, phenotypes, and genetic diagnoses [28]. This stratification of patient case complexities explains why GPT-4 performs best on the DDD dataset.

**Enhancing Diagnosis Accuracy through Multi-Agent Prompts and HPO Classification**

Given the observed variability in patient presentations and datasets during the benchmark process, we wanted to classify the patient cases by phenotype and gene-phenotype-association specificity. We used two approaches to accomplish this. We first used a multi-agent system in which an evaluator agent writes a 100-word essay assessing gene-phenotype associations and a summarizer agent analyzes the essay to classify cases as having the direct presence of gene-phenotype connections (Fig. 2A). We next used HPO Phenotype Classification methodology, which calculates the dataset specificity index (DsI) by considering HPO-term distribution across the HPO hierarchy, to reward specific terms from deeper levels of the tree and penalize general terms from levels near the root. We applied both methods to three randomly sampled sets of 90 patient cases from each of our three datasets. Together the multi-agent and HPO classification strategies allowed for a refined analysis of cases based on their phenotypic *and* genotypic characteristics.

After grouping cases using the multi-agent approach, a clear distinction emerged between cases with positive ("yes") and negative ("no) association types across the datasets (Fig. 2B, Fig. S5). There are fewer than 70 "yes" linked cases in UDN and more than 80 in the other two datasets out of 90 cases each, further confirming that UDN cases are more complex than are cases from BG or DDD. Amongst combined results from all the datasets, there is a clear distinction between the ability of the LLM to identify causal genes between either group of patient cases (Fig. 2C). For instance, the positive association cases, the LLM correctly identifies more than 30% of causal genes as the most likely gene for causing the disease but only 20% for the negative association cases.

The HPO classification method effectively differentiated the datasets into "Highly Specific HPO" and "Lowly Specific HPO" categories based on the specificity of the phenotype terms used for the patient (Fig S6). The "Highly Specific HPO" cases consistently ranked higher in terms of causal gene rank across all three datasets (BG, UDN, and DDD), indicating that more specific phenotype descriptions lead to better performance in identifying the causal genes. Furthermore, the lower performance of the LLM on the UDN cases compared to the other datasets highlights the phenotypic and genotypic diversity of patients within each dataset and emphasizes the importance of considering these factors when evaluating LLMs for patient case analysis.

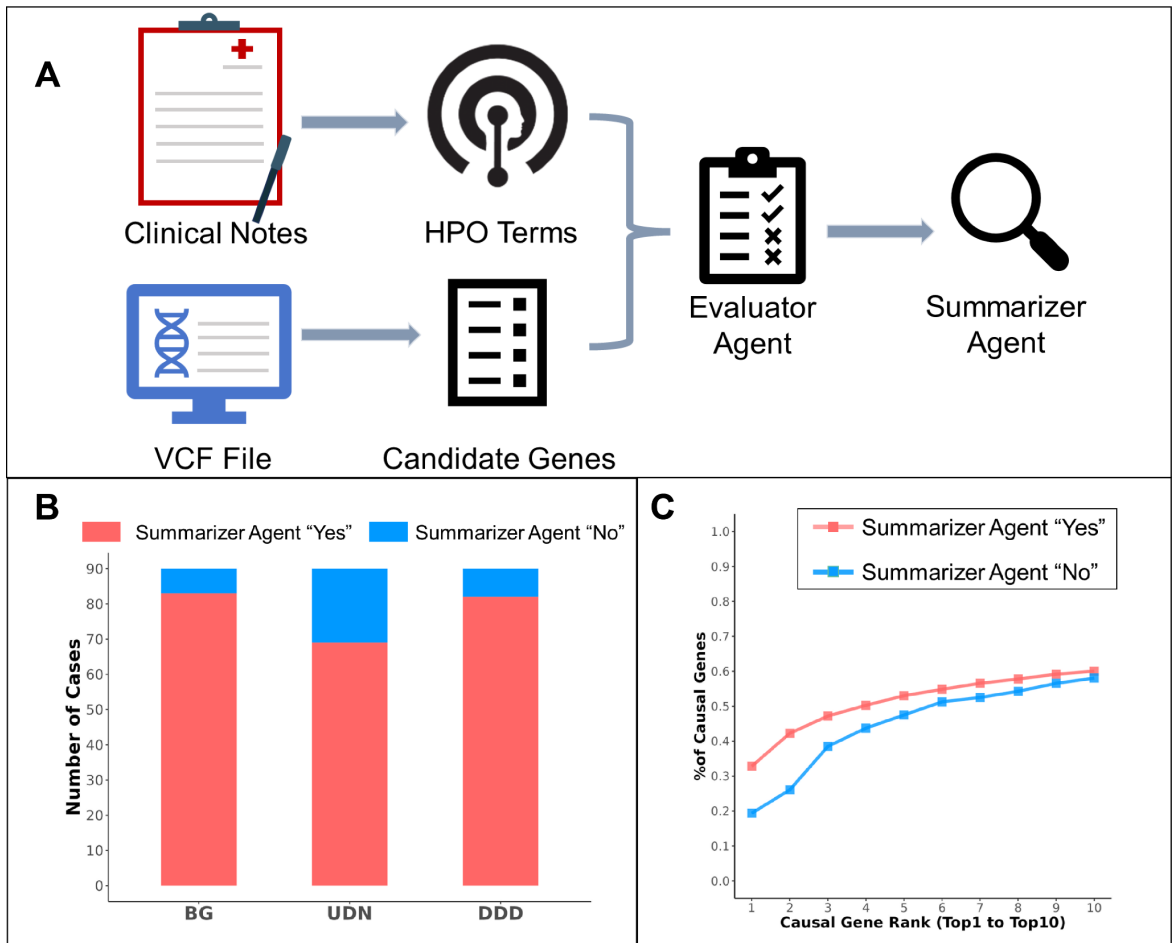

**Figure 2:** (Panel A) A schematic depicting the multi-agent approach utilizing GPT-4, including prompt creation and subsequent processing done first by the evaluator agent and subsequently summarized by the summarizer agent. (Panel B) Bar graphs showing the count of causal and non-causal genes in Baylor Genetics (BG), Undiagnosed Diseases Network (UDN), and Deciphering Developmental Disorders (DDD) databases with and without genotype-phenotype links. (Panel C) Line graphs depicting the proportion of causal genes identified by GPT-4 from ranks 1 to 10, grouped by the presence of a clear association as determined by multi-agent approach, across all datasets.

**Literature and positional bias in LLMs**

When considering the ranking abilities of LLMs it is crucial to evaluate potential biases that may arise from the representation of genes in the literature and positional biases within the prompt. To investigate these factors, we conducted an analysis of LLM performance on the BG, UDN, and DDD datasets.

Our analysis revealed a clear inverse correlation between the mean rank of a gene in these datasets and the ClinVar submission count that correspond to the gene (Fig. 3B). This finding suggests that LLMs may prioritize genes with greater representation in ClinVar, likely reflecting biases in both the literature and ClinVar itself, as well as the increased representation of these genes in the LLM pre-training data. It is well-established that genes that are more extensively studied tend to have higher submission numbers, which could explain this observed trend.

We also identified a distinct pattern in the relationship between the position of the true causal gene within the input prompt and its mean rank (Fig. 3A). As the positions of both causal genes and non-causal genes increased from 1 to 50, their mean rank also increased, indicating that the LLM assigns lower rankings to the gene when it appears later in the prompt. These observations are consistent with previous findings regarding positional bias [29,30]. Taken together, these results demonstrate that the gene ranking ability of an LLM is sensitive to the causal gene's position within the input prompt. Later positions result in higher prioritization of the true causal gene on average, underscoring the importance of accounting for such biases.

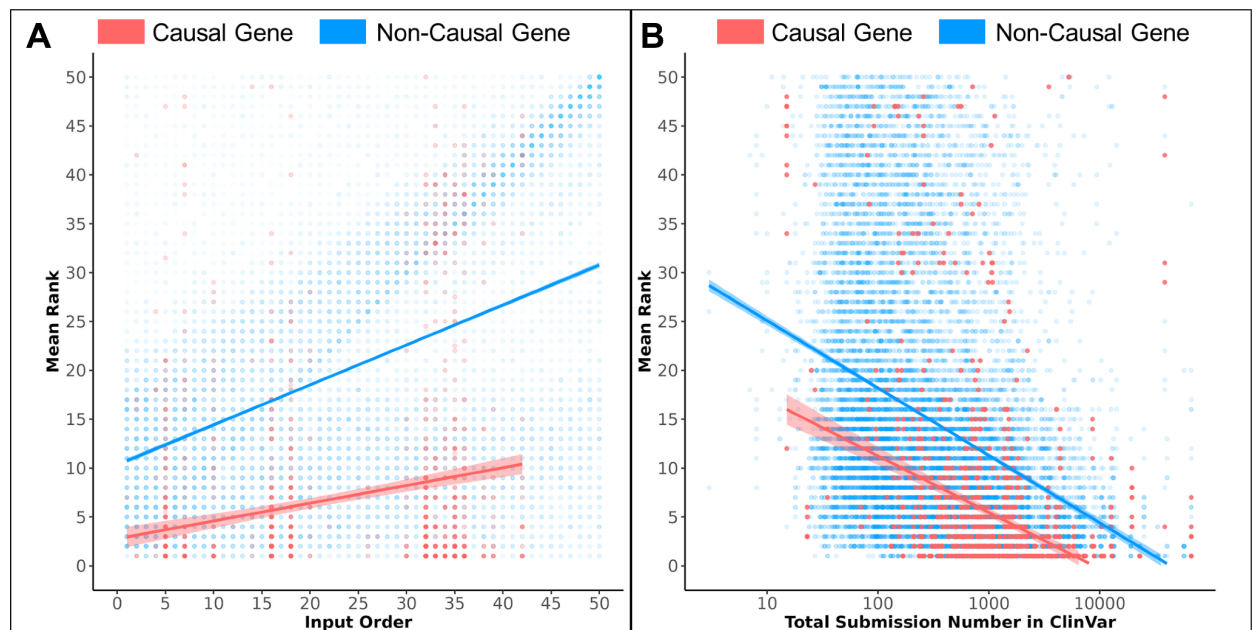

**Figure 3:** (Panel A) A scatter plot showing a positive correlation between the order of the causal gene in the prompt and mean rank shown across 50 gene positions. (Panel B) Another plot illustrates the inverse correlation between gene ranking and ClinVar submission numbers for BG, UDN, and DDD.

**Divide-and-conquer strategy addresses biases**

To counteract previously mentioned biases, we introduced the divide-and-conquer strategy to handle various gene options and orders when assigning gene rankings. This three-step pipeline involves randomly splitting gene candidates into groups of five, estimating in-group probabilities using the GPT-3.5 model, and averaging these probabilities across sampling iterations to obtain final gene scores (Fig. S7).

The divide-and-conquer strategy effectively increases the performance of GPT-3.5 for each dataset across all ranks (Fig. 4A). We observed this strategy to be more effective when implemented on ranking longer candidate gene lists (Fig. 4A, Fig. S9A). Causal genes consistently exhibit high scores when grouped with the remaining genes. Conversely, non-causal genes tend to display lower scores when grouped with causal genes, statistically scoring lower than causal genes (Fig. 4B). This strategy mitigates the literature and positional biases we have observed by employing uniform gene grouping and increasing sampling numbers. It also allows causal genes to exhibit high scores consistently while non-causal genes display lower scores when grouped with causal genes across variable lengths of candidate gene lists.

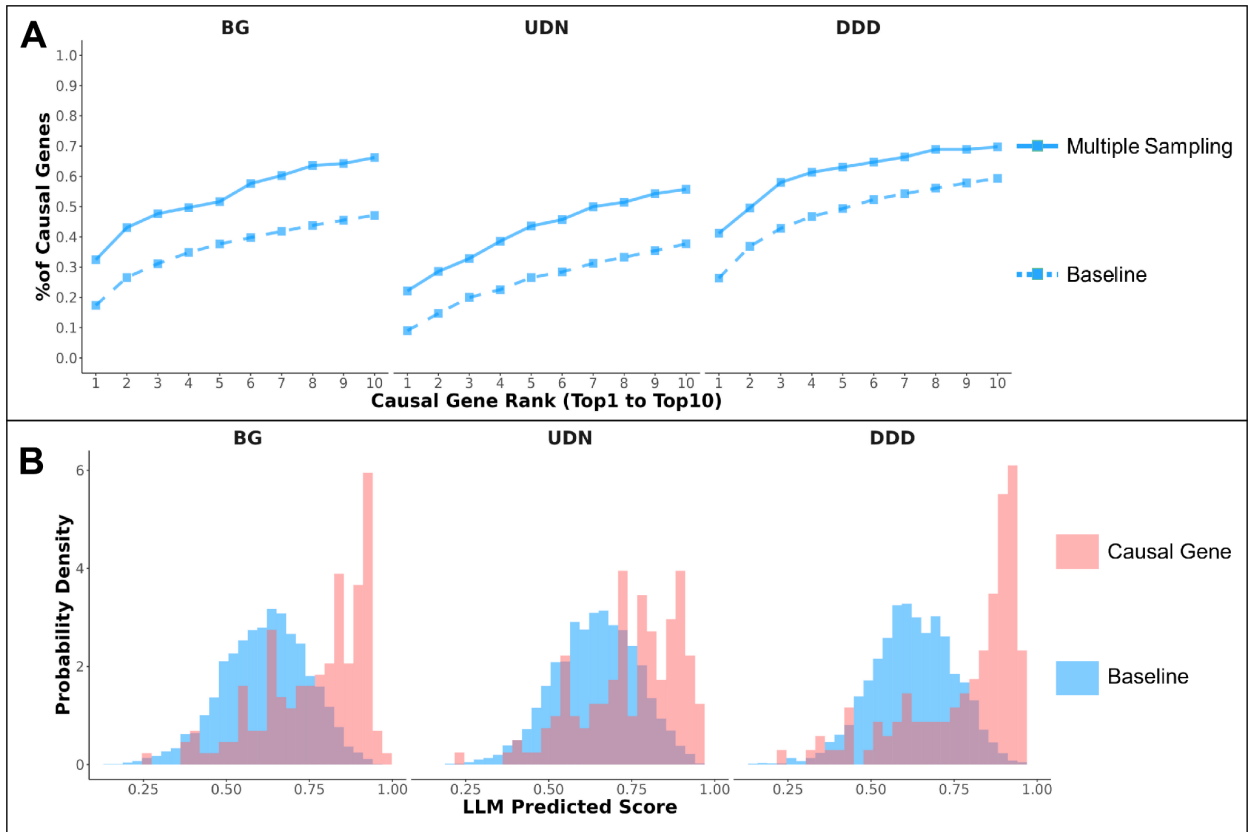

**Figure 4:** (Panel A) Line graphs show the percentage of correctly identified causal genes from ranks 1 to 10 across BG, UDN, and DDD by GPT-3.5. Solid lines indicate divide-and-conquer strategy identification rates, with dashed lines marking baseline performance.
(Panel B) Histograms display the scoring distribution of causal (blue) and non-causal (red) genes in each dataset, using the divide-and-conquer strategy.

## Discussion

Our study provides a comprehensive analysis of the ways in which Large Language Models (LLMs) can enhance gene-based prioritization, particularly in the context of rare genetic disorders. The performance benchmarks established for various LLMs, with GPT-4 as the frontrunner, underscore the potential advanced transformer architectures have to assist experts in clinical genomics. This potential is likely due to the extensive and varied training datasets that equip LLMs with an advanced understanding of both structured and unstructured medical data.

However, our findings also illuminate some challenges. Notably, certain cases were better addressed by LLMs than others, suggesting a variance in model efficacy that correlates with the complexity and specificity of the genetic and phenotypic data presented. This variability in performance emphasizes the need for models that can adapt to the high heterogeneity inherent in rare disease diagnosis.

A critical observation from our study is the apparent bias reflected in the existing literature and in genetic databases like ClinVar. Genes that are more frequently studied or reported tend to be prioritized by LLMs, potentially overshadowing less characterized but equally significant genes. This bias underscores a fundamental challenge in employing AI in medical diagnostics: the quality of AI outputs can only be as good as the data inputs.

To mitigate these biases and enhance gene-based prioritization's reliability, we propose a divide-and-conquer strategy. By dividing the prediction into multiple prioritization processes and integrating the results, our approach not only diversifies the gene candidates considered but also reduces the influence of skewed data distributions. This strategy is crucial for advancing the utility of LLMs in clinical settings, where gene prioritization accuracy can directly influence patient outcomes.

In conclusion, while LLMs offer significant promise in revolutionizing genetic diagnostics, their deployment must be carefully managed to address inherent biases and ensure equitable and comprehensive genetic analyses. Future studies should focus on refining LLMs to handle the diverse and complex nature of rare diseases more effectively, further bridging the gap between AI potential and clinical application.

## Supplementary Materials

**Output from Large Language Models**
Take text directly from GPT models as output. From open-source models, we take the probability. For the GPT models, this value is not available. We can only get text from the GPT models.

We have different methods to interact with open-source LLMs such as Llama-2 and Mixtral compared with close-source LLMs such as ChatGPT. Open-source LLMs offer the probability of each output token, a feature unavailable in closed-source LLMs like ChatGPT. To illustrate, we utilize the following prompt when querying open-source LLMs. An example is provided below, where the placeholder [GENE NAME] represents the name of a gene.

*### Given the following phenotypes in a patient: Global developmental delay, Generalized hypotonia, Failure to thrive, Tetralogy of Fallot, Hypertrophic cardiomyopathy, Abnormal facial shape. ### Are these phenotypes caused by a mutation in the [GENE NAME] gene? ### Use any available information including gene function, reports of genetic variants, expression sites, or animal model studies if direct human data is insufficient. ### Just answer Yes/No.*

The LLM outputs either 'Yes' or 'No' to indicate the causality of a gene $g_j$ to the phenotypes. We employ maximum likelihood estimation [3] to determine the causality of genes to the phenotypes. Specifically, the log-likelihood ratio of 'Yes' and 'No' indicates the log-likelihood value $S(g_j)$ that the gene $g_j$ causes the phenotypes, as given in the following equation:

$$S(g_j) = log\left(\frac{Pr(Yes)}{Pr(No)}\right),$$

where $Pr(Yes)$ and $Pr(No)$ denote the output probability of the 'Yes' and 'No' tokens. The ranking of genes is determined based on their respective log-likelihood values.

**Prompt Generation**
For HPO data, we utilized the get_ontology() function from the ontologyIndex package to retrieve the HPO data. For gene data, we first extracted the gene symbols of causal variants from every patient case. Then, we used a sampling with pre-defined seeds to sample *n-k* gene symbols of non-causal variants from the filtered gene pool (*n* is the pre-defined number for all input genes and *k* is the number of gene symbols of causal variants of the given case). After the sampling, the causal genes and non-causal genes of every case will be combined in a shuffled order to reduce the positional bias. For investigating the effect of input order on the rank of causal genes, 5 different seeds were used in the shuffling process to get varied input orders.

**GPT Query**
Queries of GPT-3.5 and GPT-4 are performed by calling API. The model used for GPT-3.5 is gpt-3.5-turbo-1106. The model used for GPT-4 is gpt-4-1106-preview. The temperature is set to 1.0. For every patient case, 5 repeated queries with the same prompt will be performed for every case to evaluate the baseline robustness of the GPT models. Based on the output of API calling, another GPT-3.5 agent (gpt-3.5-turbo-1106, temperature = 1.0) will check and summarize the result into a list with the format format: Gene Name - Probability to reduce inconsistency of the output formats. Because of the consumption of computational resources and the cost of API calls, we only did one query instead of 5 repeated queries for the multi-agent approach and correlation of input order and rank.

**Result Summary**
Based on the summarized output of LLMs, we matched the output gene names with the input gene names in the same case to assign ranks to every gene in the given case. Handling hallucinations, if the causal gene in the input list is not in the output list, the case will be abandoned for a further summary. Based on our observation, hallucinations with irrelevant outputs are rare. However, the output usually contains fewer genes compared with the input gene lists, especially when the input number of genes is 25 and 50. To measure the proportion of genes that can be matched during the result summary, we have causal_gene_output_ratio in the supplementary tables to evaluate the proportion of causal genes that can be matched in every batch of query and match_mean to evaluate the mean proportion of all input genes that can be matched in every batch of query. After assigning ranks to every gene, the metrics "% of causal genes" will be calculated by calculating the proportion of causal genes ranked ≤ n among all the causal genes in the batch of queries. Therefore, the abandoned cases due to missing causal genes will still be penalized as not within the given rank number.

**Decreasing Inference/Generation Latency**
To run larger open-source LLMs, we employ mini-batch inference and output clipping technologies. Specifically, mini-batch inference is a technique of feeding mini-batches of the input sequence to LLMs, thereby accelerating the overall generation process by distributing the inference processes. Output clipping is a technique utilized to reduce inference time by limiting the output token number, thereby shortening the output sequence. This approach is simply implemented by setting the `max_new_tokens` argument parameter to a low value. It proves particularly effective in tasks where concise answers are sufficient, such as confirming gene-disease associations with simple affirmative or negative responses. The reduction in output tokens can effectively reduce the latency without compromising performance.

| Model | Gene Number | Dataset | Top1 Mean | Top1 SD | Top3 Mean | Top3 SD | Top5 Mean | Top5 SD | Top10 Mean | Top10 SD | Casual Gene Output Ratio | Total Gene Output Ratio |
|---|---|---|---|---|---|---|---|---|---|---|---|---|
| GPT3.5 | 5 | BG | 0.42 | 0.00 | 0.75 | 0.00 | 1.00 | 0.00 | 1.00 | 0.00 | 1.00 | 1.00 |
| GPT3.5 | 5 | UDN | 0.40 | 0.03 | 0.75 | 0.01 | 0.99 | 0.01 | 0.99 | 0.01 | 0.99 | 1.00 |
| GPT3.5 | 5 | DDD | 0.60 | 0.01 | 0.83 | 0.01 | 1.00 | 0.00 | 1.00 | 0.00 | 1.00 | 1.00 |
| GPT3.5 | 25 | BG | 0.32 | 0.01 | 0.51 | 0.00 | 0.58 | 0.00 | 0.70 | 0.00 | 0.91 | 0.85 |
| GPT3.5 | 25 | UDN | 0.25 | 0.02 | 0.42 | 0.02 | 0.51 | 0.01 | 0.66 | 0.02 | 0.87 | 0.82 |
| GPT3.5 | 25 | DDD | 0.46 | 0.01 | 0.62 | 0.01 | 0.69 | 0.01 | 0.78 | 0.02 | 0.95 | 0.86 |
| GPT3.5 | 50 | BG | 0.17 | 0.00 | 0.31 | 0.01 | 0.38 | 0.01 | 0.47 | 0.01 | 0.59 | 0.36 |
| GPT3.5 | 50 | UDN | 0.09 | 0.01 | 0.20 | 0.02 | 0.27 | 0.02 | 0.38 | 0.05 | 0.54 | 0.39 |
| GPT3.5 | 50 | DDD | 0.26 | 0.02 | 0.43 | 0.01 | 0.49 | 0.02 | 0.59 | 0.01 | 0.69 | 0.36 |
| GPT4 | 5 | BG | 0.53 | 0.00 | 0.78 | 0.00 | 1.00 | 0.00 | 1.00 | 0.00 | 1.00 | 1.00 |
| GPT4 | 5 | UDN | 0.55 | 0.01 | 0.77 | 0.01 | 1.00 | 0.00 | 1.00 | 0.00 | 1.00 | 1.00 |
| GPT4 | 5 | DDD | 0.72 | 0.00 | 0.83 | 0.01 | 1.00 | 0.00 | 1.00 | 0.00 | 1.00 | 1.00 |
| GPT4 | 25 | BG | 0.43 | 0.00 | 0.60 | 0.01 | 0.66 | 0.00 | 0.74 | 0.00 | 0.98 | 0.96 |
| GPT4 | 25 | UDN | 0.32 | 0.01 | 0.52 | 0.01 | 0.62 | 0.02 | 0.75 | 0.03 | 0.94 | 0.96 |
| GPT4 | 25 | DDD | 0.63 | 0.00 | 0.73 | 0.00 | 0.76 | 0.01 | 0.81 | 0.01 | 0.99 | 0.98 |
| GPT4 | 50 | BG | 0.30 | 0.00 | 0.43 | 0.00 | 0.48 | 0.01 | 0.56 | 0.00 | 0.76 | 0.60 |
| GPT4 | 50 | UDN | 0.19 | 0.01 | 0.29 | 0.04 | 0.40 | 0.02 | 0.53 | 0.01 | 0.67 | 0.52 |
| GPT4 | 50 | DDD | 0.44 | 0.01 | 0.61 | 0.01 | 0.64 | 0.01 | 0.69 | 0.02 | 0.82 | 0.60 |

**Supplementary Table S1:** Benchmark results of the OpenAI models GPT-4 and GPT-3.5-turbo averaged across 5 repetitions of LLM interactions. The table includes the number of candidate genes to be ranked by the LLMs, the casual gene output ratio represents the average ratio of causal genes in the output to the total number of causal genes across all cases, representing the LLMs ability to include the causal gene in the ranking. The total gene output ratio is the proportion of genes in the input that are still represented in the output. The Top*N* columns represent the proportion of causal genes that were ranked at or within the ranking position *N*.

| Model | Sample Size | Dataset | Top1 Mean | Top3 Mean | Top5 Mean | Top10 Mean |
| --- | --- | --- | --- | --- | --- | --- |
| Llama-2-70B | 5.00 | BG | 0.32 | 0.67 | 1.00 | 1.00 |
| Llama-2-70B | 5.00 | UDN | 0.34 | 0.70 | 1.00 | 1.00 |
| Llama-2-70B | 5.00 | DDD | 0.42 | 0.72 | 1.00 | 1.00 |
| Llama-2-70B | 25.00 | BG | 0.15 | 0.29 | 0.39 | 0.56 |
| Llama-2-70B | 25.00 | UDN | 0.14 | 0.24 | 0.38 | 0.57 |
| Llama-2-70B | 25.00 | DDD | 0.20 | 0.32 | 0.44 | 0.61 |
| Llama-2-70B | 50.00 | BG | 0.10 | 0.19 | 0.25 | 0.38 |
| Llama-2-70B | 50.00 | UDN | 0.09 | 0.15 | 0.21 | 0.38 |
| Llama-2-70B | 50.00 | DDD | 0.11 | 0.22 | 0.30 | 0.44 |
| Mixtral-8x7B | 5.00 | BG | 0.44 | 0.76 | 1.00 | 1.00 |
| Mixtral-8x7B | 5.00 | UDN | 0.44 | 0.74 | 1.00 | 1.00 |
| Mixtral-8x7B | 5.00 | DDD | 0.58 | 0.81 | 1.00 | 1.00 |
| Mixtral-8x7B | 25.00 | BG | 0.25 | 0.41 | 0.50 | 0.65 |
| Mixtral-8x7B | 25.00 | UDN | 0.19 | 0.38 | 0.51 | 0.69 |
| Mixtral-8x7B | 25.00 | DDD | 0.36 | 0.57 | 0.63 | 0.75 |
| Mixtral-8x7B | 50.00 | BG | 0.20 | 0.32 | 0.39 | 0.52 |
| Mixtral-8x7B | 50.00 | UDN | 0.13 | 0.28 | 0.39 | 0.51 |
| Mixtral-8x7B | 50.00 | DDD | 0.24 | 0.47 | 0.54 | 0.64 |
| BioMistral-7B | 5.00 | BG | 0.35 | 0.72 | 1.00 | 1.00 |
| BioMistral-7B | 5.00 | UDN | 0.33 | 0.71 | 1.00 | 1.00 |
| BioMistral-7B | 5.00 | DDD | 0.45 | 0.80 | 1.00 | 1.00 |
| BioMistral-7B | 25.00 | BG | 0.17 | 0.32 | 0.41 | 0.60 |
| BioMistral-7B | 25.00 | UDN | 0.15 | 0.30 | 0.35 | 0.60 |
| BioMistral-7B | 25.00 | DDD | 0.18 | 0.35 | 0.47 | 0.64 |
| BioMistral-7B | 50.00 | BG | 0.12 | 0.22 | 0.28 | 0.41 |
| BioMistral-7B | 50.00 | UDN | 0.10 | 0.14 | 0.23 | 0.34 |
| BioMistral-7B | 50.00 | DDD | 0.11 | 0.23 | 0.31 | 0.44 |

**Supplementary Table S2:** Benchmark results of the open-source LLMs. The table includes the number of candidate genes to be ranked by the LLMs. The Top*N* columns represent the proportion of causal genes that were ranked at or within the ranking position *N*.

**A**

Given the following phenotypes in a patient:
Duplicated collecting system
Global developmental delay
Generalized hypotonia
Failure to thrive
Bicuspid aortic valve
Abnormal facial shape

And considering these genes:
CASD1
NEO1
PASD1
ZNF560
KMT2D

Task: Rank the genes by their likelihood of causing the patient's phenotypes. Use any available information including gene function, reports of genetic variants, expression sites, or animal model studies if direct human data is insufficient.

Format: Provide a ranked probability (0.00 to 0.99) list with two columns (the name of the gene and the probability) without any introductory sentences and without any explanations. The format of every row of the ranked list should be: Gene Name – Probability.

**B**

Given the following phenotypes in a patient:
### Given the following phenotypes in a patient: Global developmental delay, Generalized hypotonia, Failure to thrive, Tetralogy of Fallot, Hypertrophic cardiomyopathy, Abnormal facial shape. ### Are these phenotypes caused by a mutation in the [GENE NAME] gene? ### Use any available information including gene function, reports of genetic variants, expression sites, or animal model studies if direct human data is insufficient. ### Just answer Yes/No.

**Supplementary Figure S1:** (Panel A) An example of the prompt for GPT 3.5 and GPT 4. Panel B) An example of the prompt for open-source LLMs.

Given the phenotypes: Lymphopenia, … Are these caused by the MRPS10P2 gene? Yes/No
Given the phenotypes: Visual loss, … Are these caused by the RPL7P42 gene? Yes/No
Given the phenotypes: Dysphagia, … Are these caused by the ARSA gene? Yes/No
Given the phenotypes: Arthralgia, … Are these caused by the DDX11 gene? Yes/No
Given the phenotypes: Delayed speech … Are these caused by the WDR19 gene? Yes/No
Given the phenotypes: Intellectual disability … Are these caused by the ATP1B1 gene? Yes/No
Given the phenotypes: Abnormality … Are these caused by the ZNF645 gene? Yes/No
Given the phenotypes: Hydrocephalus, … Are these caused by the ATP1B1 gene? Yes/No

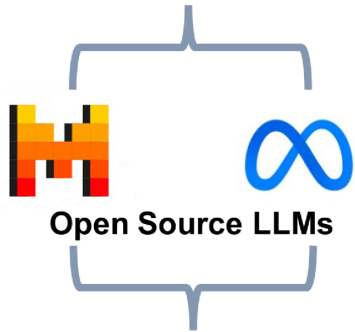

**Open Source LLMs**

P(Yes) = 0.6; P(No) = 0.4
P(Yes) = 0.8; P(No) = 0.2
P(Yes) = 0.3; P(No) = 0.7
P(Yes) = 0.55; P(No) = 0.45
P(Yes) = 0.75; P(No) = 0.25
P(Yes) = 0.7; P(No) = 0.3
P(Yes) = 0.25; P(No) = 0.75
P(Yes) = 0.15; P(No) = 0.85

Estimating Log-likelihood Value

| Gene | Log-likelihood |
|---|---|
| MRPS10P2 | log(0.6/0.4) |
| RPL7P42 | log(0.8/0.2) |
| ARSA | log(0.3/0.7) |
| DDX11 | log(0.55/0.45) |
| WDR19 | log(0.75/0.25) |
| ATP1B1 | log(0.7/0.3) |
| ZNF645 | log(0.25/0.75) |
| ATP1B1 | log(0.15/0.85) |

**Supplementary Figure S2:** An example of using mini-batch inference and log-likelihood estimation in open-source LLMs.

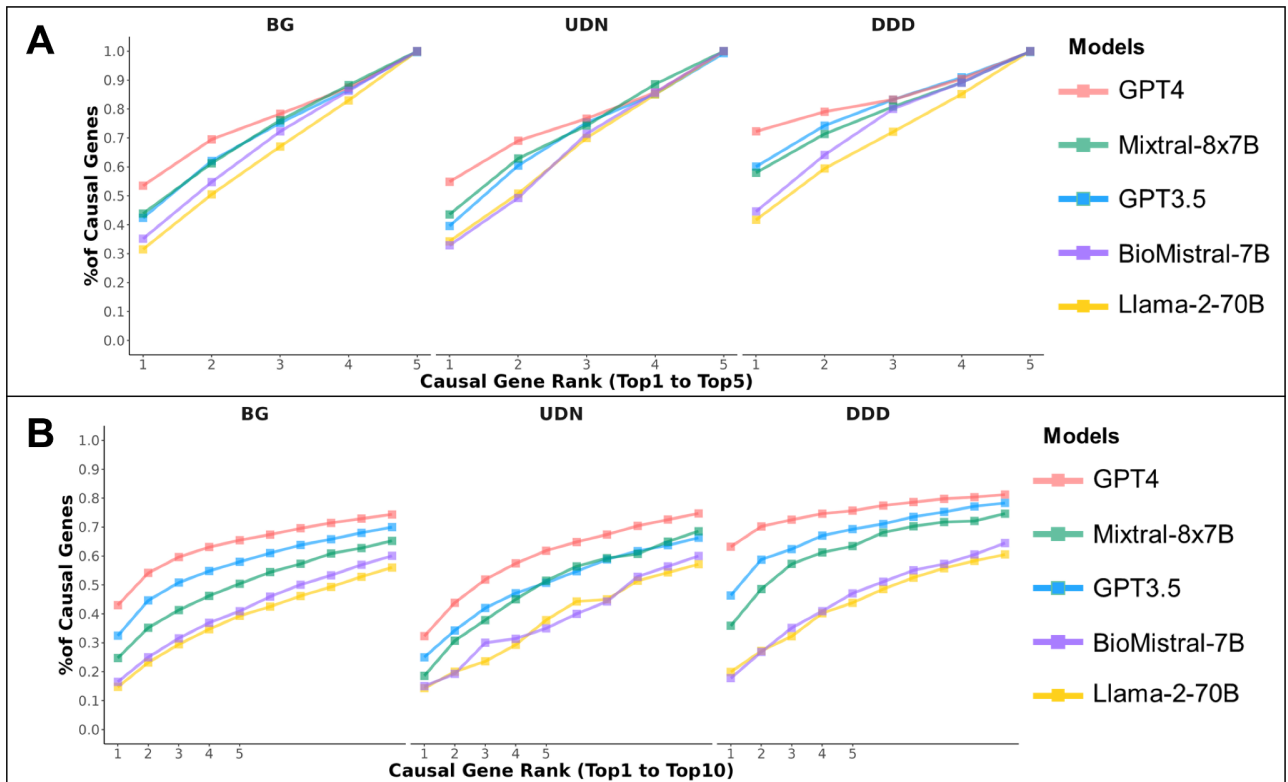

**Supplementary Figure S3:** (Panel A) Comparison of LLMs performance in ranking causal genes with 5 input genes per case. (Panel B) Comparison of LLMs performance in ranking causal genes with 25 input genes per case.

**Supplementary Figure S4:** (Panel A) An example of the prompt for the Evaluator Agent. (Panel B) An example of the prompt for the Summarizer Agent.

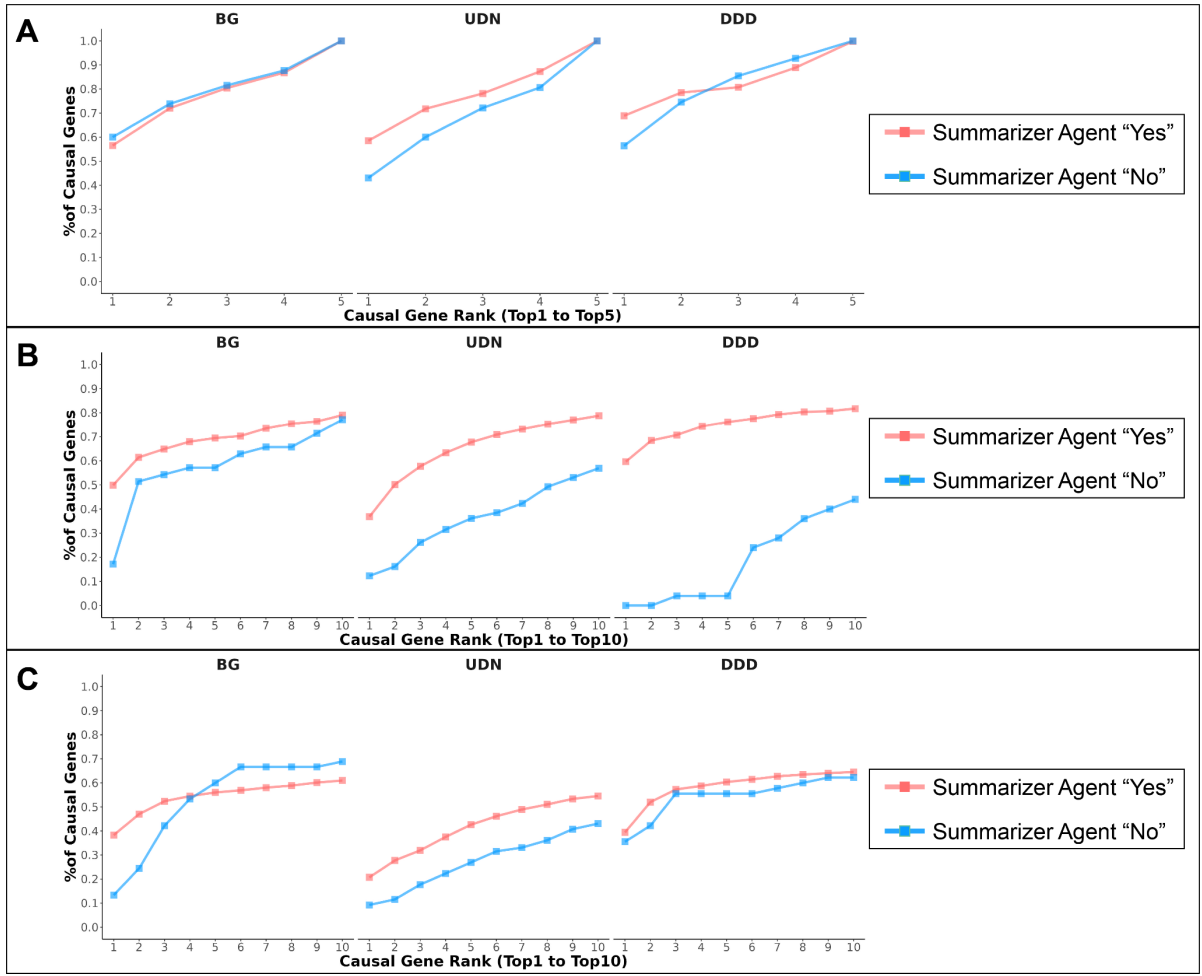

**Supplementary Figure S5:** (Panel A) Line graphs depicting the proportion of causal genes identified from ranks 1 to 10 under the multi-agent approach with 5 input genes per case. (Panel B) Line graphs depicting the proportion of causal genes identified from ranks 1 to 10 under the multi-agent approach with 25 input genes per case. (Panel C) Line graphs depicting the proportion of causal genes identified from ranks 1 to 10 under the multi-agent approach with 50 input genes per case.

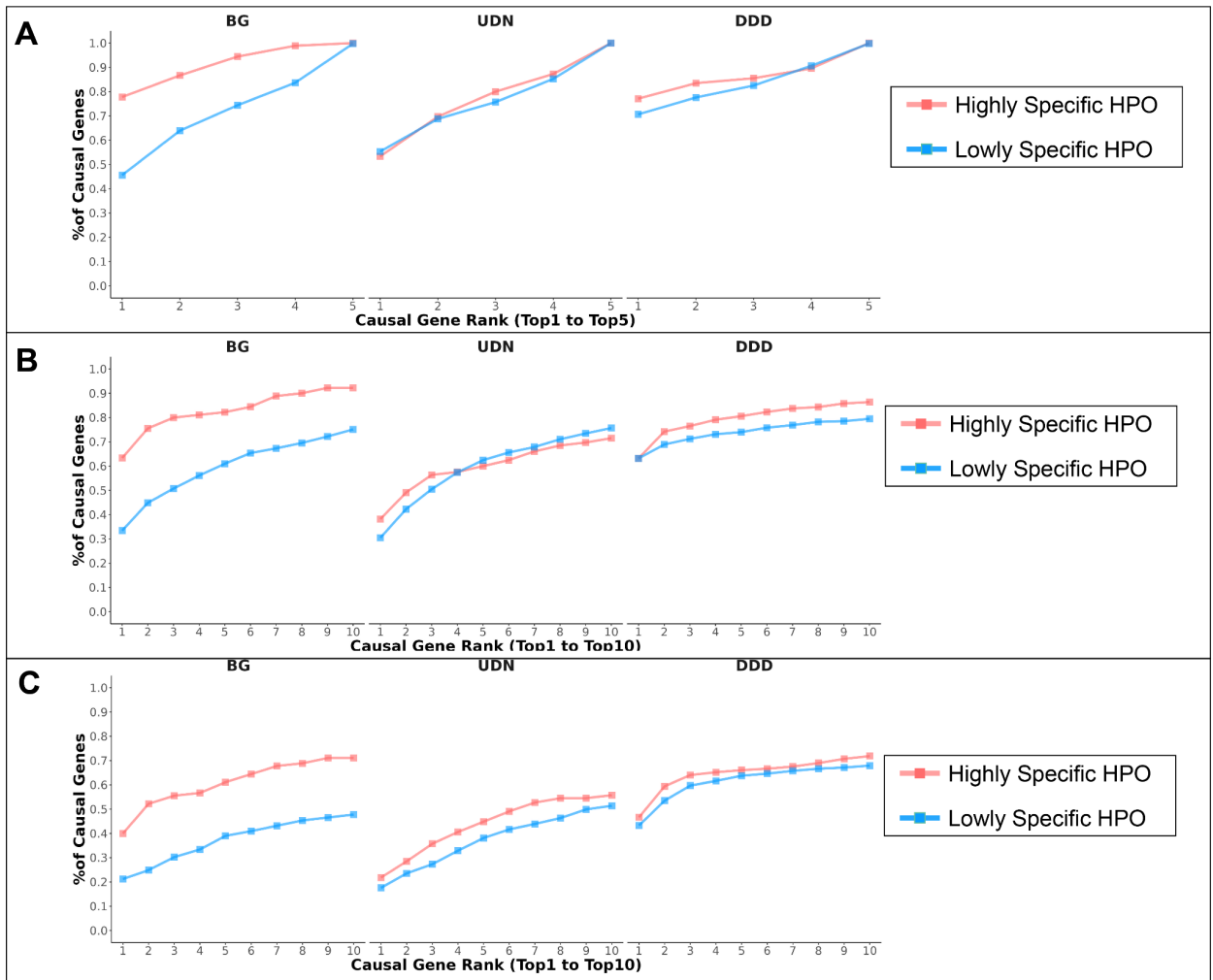

**Supplementary Figure S6:** (Panel A) Line graphs depicting the proportion of causal genes identified from ranks 1 to 10 under the HPO-classification approach with 5 input genes per case. (Panel B) Line graphs depicting the proportion of causal genes identified from ranks 1 to 10 under the HPO-classification approach with 25 input genes per case. (Panel C) Line graphs depicting the proportion of causal genes identified from ranks 1 to 10 under the HPO-classification approach with 50 input genes per case.

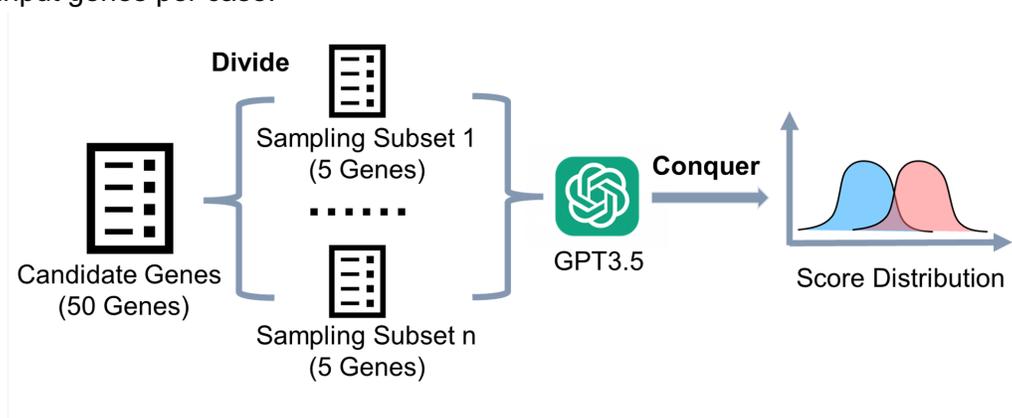

**Supplementary Figure S7:** The workflow of implementing the Divide-Conquer Approach

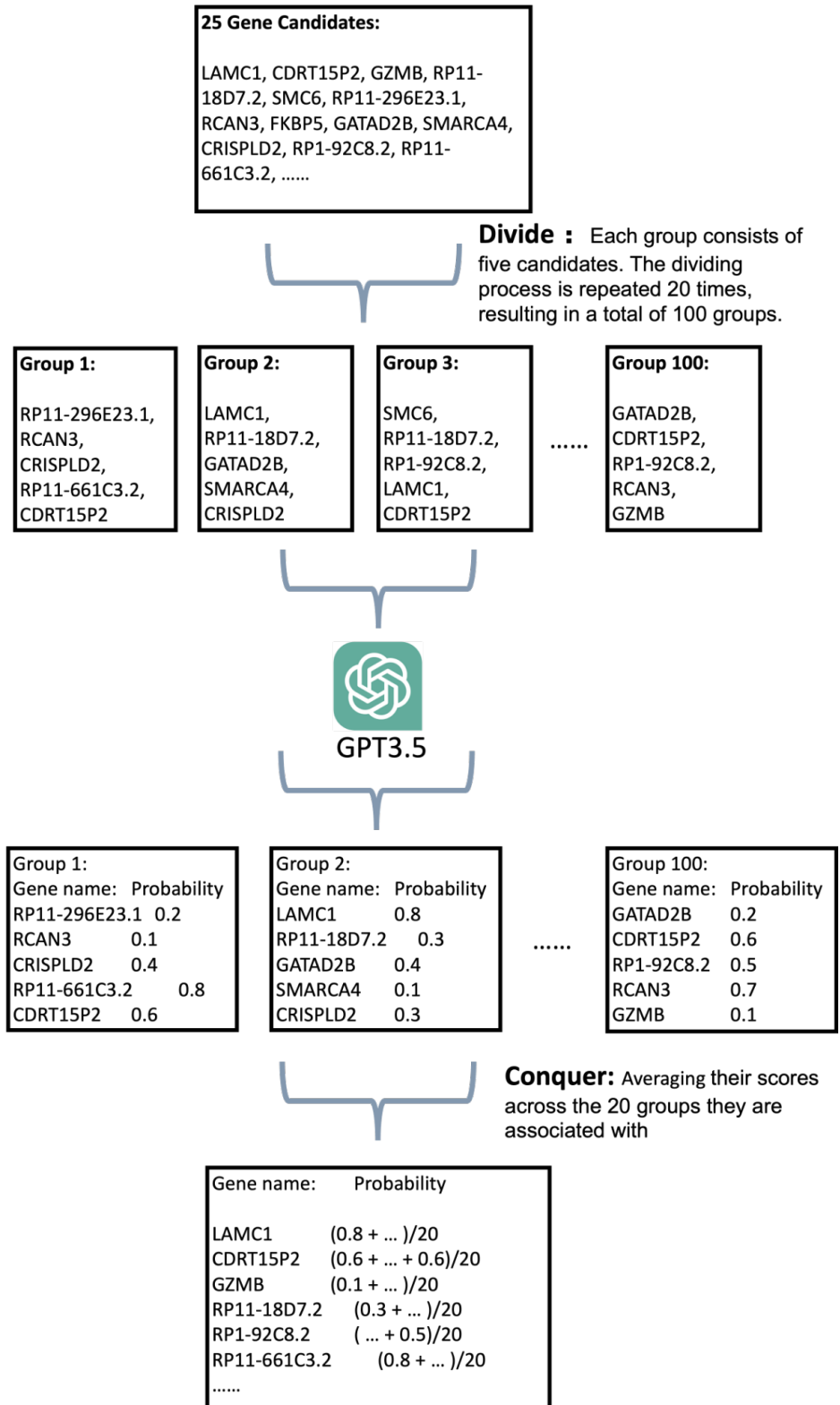

**Supplementary Figure S8:** An example of the Divide-and-conquer Approach. Let's consider a scenario where we have 25 gene candidates that need to be ranked. We begin by dividing

these candidates into groups, each consisting of five candidates. This process is repeated 20 times, resulting in a total of 100 groups. Consequently, every gene is part of 20 distinct groups. Subsequently, we employ Figure S1 to estimate the ranking of genes within each group. Finally, we compute the average of their scores across the 20 groups they are associated with.

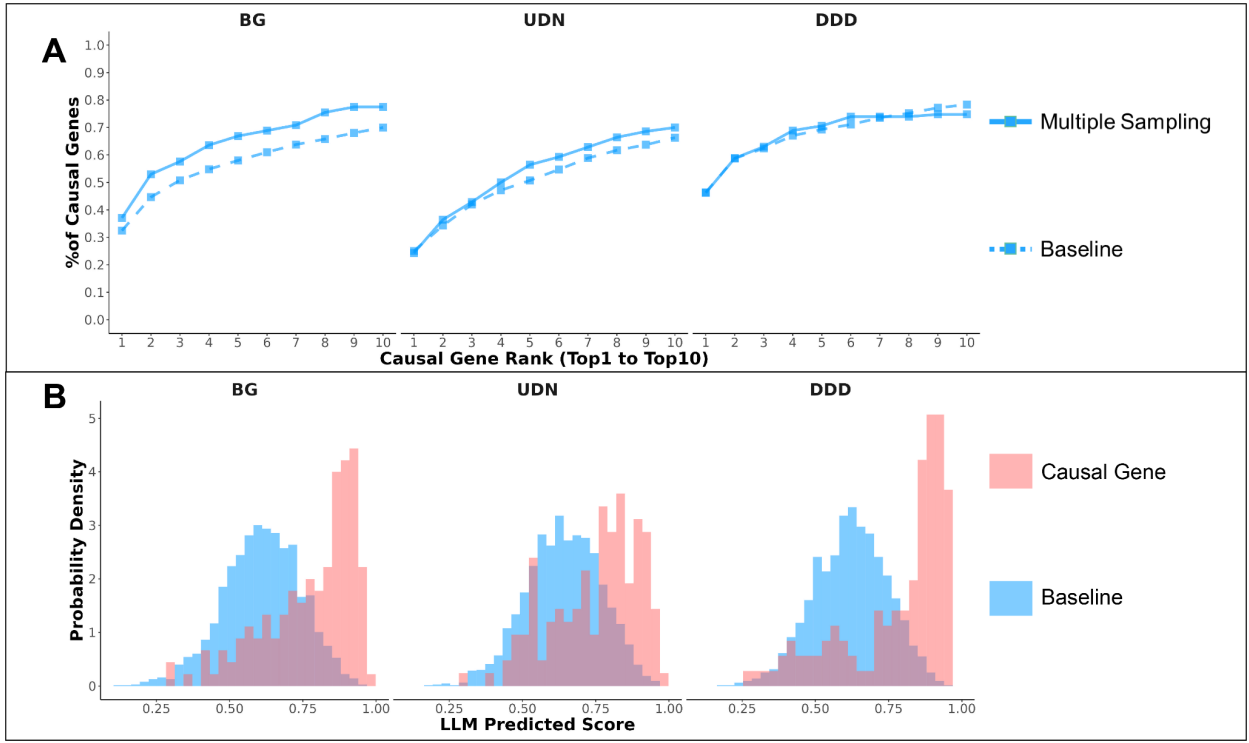

**Supplementary Figure S9:** (Panel A) Line graphs show the percentage of correctly identified causal genes from ranks 1 to 10 across BG, UDN, and DDD by GPT-3.5. Solid lines indicate divide-and-conquer strategy identification rates with 25 input genes per case, with dashed lines marking baseline performance. (Panel B) Histograms display the scoring distribution of causal (blue) and non-causal (red) genes in each dataset, using the divide-and-conquer strategy with 25 input genes per case.

**A**

**USER**

What is the corresponding definition for the human phenotype ontology term HP:0009752?

**ASSISTANT**

The corresponding definition for the human phenotype ontology term HP:0009752 is "Narrow palate".

**B**

**USER**

What is the corresponding definition for the human phenotype ontology term HP:0030036?

**ASSISTANT**

The corresponding definition for the human phenotype ontology term HP:0030036 is "Abnormality of the immune system."

**Supplementary Figure S10:** (A) An interaction with GPT-3.5-turbo in the OpenAI playground with default settings beside temperature set to 0.5. Queried on the HPO ID HP:0030036 where the correct definition is "Isothenuria" the LLM returned a hallucination which we have observed is common behavior for LLMs in regards to understanding HPO IDs. (B) Another hallucination from the same LLM and parameters for the HPO term and ID "Cleft in skull base" HP:0009752.